\documentclass[nofootinbib,aps,superscriptaddress,11pt]{revtex4-2}
\usepackage{geometry}
\usepackage{graphicx}
\usepackage{dcolumn}
\usepackage{bm}
\usepackage{epsfig,amsmath}
\usepackage{amssymb}
\usepackage{subfigure,esint}
\usepackage{float}
\usepackage[usenames,dvipsnames]{color}
\usepackage[pagebackref=false, colorlinks=true]{hyperref}
\definecolor{redish}{rgb}{0.7,0.2,0.0}  
\definecolor{bluish}{rgb}{0.2,0.5,0.8}

\hypersetup{linkcolor=redish,          
                  citecolor=blue,        
                  filecolor=magenta,      
                  urlcolor=blue}          

\begin{document}
\author{Rajibul Shaikh}\email{rshaikh@iitk.ac.in}
\affiliation{Department of Physics, Indian Institute of Technology, Kanpur 208016, India}
\author{Suvankar Paul}\email{svnkr@iitk.ac.in}
\affiliation{Department of Physics, Indian Institute of Technology, Kanpur 208016, India}
\affiliation{Department of Physics, ICFAI University, Agartala, Tripura 799210, India}
\author{Pritam Banerjee}\email{bpritam@iitk.ac.in}
\affiliation{Department of Physics, Indian Institute of Technology, Kanpur 208016, India}
\author{Tapobrata Sarkar}
\email{tapo@iitk.ac.in}
\affiliation{Department of Physics, Indian Institute of Technology, Kanpur 208016, India}

\title{\Large Shadows and thin accretion disk images of the $\gamma$-metric}

\bigskip

\begin{abstract}
The $\gamma$-metric is a static, axially-symmetric
singular solution of the vacuum Einstein's equations without an event horizon. 
This is a two-parameter family of solutions, generic values of 
one of which (called $\gamma$) measure the deviation from spherical symmetry. Here, we 
first study the shadow cast by this geometry, in order to constrain the $\gamma$-metric from observations. 
We find that for $\gamma < 1/2$, there 
are, in principle, no shadows cast. On the other hand, shadows cast for all values of $\gamma \geq 1/2$ are consistent 
with observations of M$87^*$ by the Event Horizon Telescope. We also study images of 
thin accretion disks in the $\gamma$-metric background. In situations where the $\gamma$-metric possesses light rings, these qualitatively mimic Schwarzschild black holes with the same ADM mass, while in the absence of such rings, 
they are drastically different from the black hole case. 

\end{abstract}
\maketitle


\section{Introduction}

Physics near the event horizon of a black hole has been one of the most fascinating topics of research ever since
the discovery of General Relativity (GR) more than a century ago. Indeed, a singularity indicates the limits
of applicability of a theory, and the event horizon, which cloaks the central singularity of a black hole 
might hold the key to understanding the most fundamental aspects of gravity. These include the elusive
quantum aspects of gravity, which, many believe, should smoothen this singularity, which is often an 
inevitable end-state of gravitational collapse in GR. 

Studies related to singularities and event horizons assume extreme importance and relevance 
given the fact that it is now commonly believed that the centers of most galaxies are inhabited by 
supermassive black holes. While such studies were purely of theoretical interest till a few years back, the advent of the Event Horizon Telescope (EHT)
has proved to be a game changer. The phenomenal advances in observational studies near the event horizon of
black holes has ushered in a new era where one can meaningfully compare theoretical results with EHT data. 
In fact, immediately after the EHT collaboration announced its first results on the radio source M$87^*$ \cite{EHT1,EHT2,EHT3}, a flurry of activities have started, one of the most important being the constraining of various solutions of GR and many other gravity theories that are either black holes or can mimic one. The standard way to do this is to compare the theoretically obtained
shadow with the one reported by the EHT for M$87^*$. Indeed, this has been shown to put stringent constraints
on the parameters of the underlying theory \cite{tapo1,Psaltis,Soumitra,Bambi,Rahul}.

In astrophysical scenarios, the spacetime geometry around a supermassive compact object is typically modeled by a static Schwarzschild or a stationary
Kerr black hole solution of GR. While Birkhoff's theorem guarantees that the former is the unique solution under the assumption
of spherical symmetry in vacuum, the latter is perhaps more interesting as a rotating solution. Such a solution
is axially symmetric, and thus more general than the idealized Schwarzschild solution. While the Kerr black hole
has been extensively studied in the light of EHT data, a couple of its mimickers were recently studied in the
same context in \cite{tapo1}, where it was pointed out that such alternatives are indeed viable, within some 
particular range of parameters. 

In astrophysical scenarios, axially symmetric solutions of Einstein's equations are interesting, as 
these are more general than idealized spherically symmetric cases. In the simplest scenario, one can
envisage a static solution, which can have higher multipole terms in its potential, apart from the usual monopole term. Properties of 
such spacetimes are well studied in the literature, and exact solutions are known 
\cite{Erez-Rosen_1959,Zipoy_1966,Voorhees_1970,Esposito_1975,Gutsunaev_1985} (for more details, we
refer the reader to the monographs \cite{Stephani} and \cite{Griffiths_2009}). Here, we will focus on the
Zipoy-Voorhees spacetime \cite{Zipoy_1966,Voorhees_1970,Stephani,Griffiths_2009,Herrera_1999,Kodama_2003}, whose
metric is popularly known as the $\gamma$-metric. These are singular and vacuum solutions of Einstein's
equations without an event horizon and
are characterized by two parameters, which we call $m$ and $\gamma$. While $m$ relates to the mass, 
$\gamma$ has two special values, namely $\gamma=0$ representing flat space and $\gamma = 1$ being the Schwarzschild
solution. For all other values of $\gamma$, the space-time is axially symmetric, with $\gamma$ being 
a measure of the deviation from spherical symmetry or from the Schwarzschild
solution. 

Our purpose here is to check how far the $\gamma$-metric can act as a black hole mimicker. To this end, we first
study the shadow cast by the $\gamma$-metric, and compare it with the EHT result. Our main observation
is that while for $\gamma < 1/2$, there will in principle be no shadow (this contradicts the recent
results of \cite{Galtsov_2019,Bambi_2019}), while the EHT data indicates that 
{\it all values} of $\gamma \geq 1/2$ are allowed. This 
is verified by studying the gamma metric in the limit $\gamma \to \infty$, where it coincides with the
Chazy-Curzon solution in spherical coordinates \cite{Griffiths_2009}. We, therefore, come up with an
interesting conclusion : the gamma metric, with $\gamma \geq 1/2$ is an {\it unconstrained} axially symmetric 
solution of vacuum GR, consistent with current EHT data. In this sense, the shadow of the 
$\gamma$-metric adds to the list of black hole mimickers. 

In the later part of this paper, we focus on geometrically thin accretion disk images for the $\gamma$-metric. We consider
the Novikov-Thorne model, and study massive particles on the ``equatorial'' plane. The images of the
disks are then numerically obtained using ray-tracing technique and compared with those of Schwarzschild black hole. Our conclusion here is that, in the presence of light rings, the 
accretion disk images of the $\gamma$-metric can very closely mimic those of the Schwarzschild black hole, but in the
absence of such rings it is drastically different. 

This paper is organized as follows. In the next section \ref{section2}, we review some properties of the $\gamma$-metric,
and also the metric in the limit $\gamma \to \infty$, in subsection \ref{limiting}. Next, in section \ref{shadow}, we
study the shadow cast by the $\gamma$-metric. This is followed by the section \ref{constrain}, where we 
constrain the $\gamma$-metric using EHT results, substantiating our discussion above. Section $\ref{accretion}$
is devoted to the analysis of thin accretion disks in the $\gamma$-metric background, followed by the 
concluding section \ref{conclusions}. Throughout this paper, we will work in units where the gravitational
constant $G$ and the speed of light $c$ is set to unity. 

\section{The $\gamma$-metric and its properties}
\label{section2}

We consider a particular solution of the Weyl class which consists of a family of static, axially symmetric vacuum 
solutions of Einstein's equations. The solution is called as the Zipoy-Voorhees spacetime 
\cite{Zipoy_1966,Voorhees_1970,Stephani,Griffiths_2009,Herrera_1999,Kodama_2003}, whose metric is popularly known as 
the $\gamma$-metric. The Weyl class of metrics have a generic line element of the form \cite{Stephani,Griffiths_2009,Herrera_1999}
\begin{equation}
ds^2 = -e^{2U}dt^2 +e^{-2U}\left(e^{2k}(d\rho^2+dz^2) + \rho^2d\phi^2\right),
\label{Weyl}
\end{equation}
in cylindrical coordinates $(t,\rho,\phi,z)$. The $\gamma$-metric, for a particular solution of $U(\rho,z)$, $k(\rho,z)$ and after making a transformation from $(\rho,z)$ to $(r,\theta)$ coordinates, is written as 
\begin{align}
\nonumber ds^2 &= g_{tt}dt^2+g_{rr}dr^2+g_{\theta\theta}d\theta^2+g_{\phi\phi}d\phi^2 \\
 &= -A(r)dt^2 + \frac{1}{A(r)}\left[ B(r,\theta)dr^2 + C(r,\theta) d\theta^2 + (r^2-2mr)\sin^2\theta d\phi^2 \right],
\label{eq:metric_gamma}
\end{align}
where the functions $ A,B,C $ are given as
\begin{equation}
A(r) = \left( 1-\frac{2m}{r} \right)^\gamma,~B(r,\theta) = \left( \frac{r^2-2mr}{r^2-2mr+m^2\sin^2\theta} \right)^{\gamma^2-1},~
C(r,\theta) = \frac{(r^2-2mr)^{\gamma^2}}{(r^2-2mr+m^2\sin^2\theta)^{\gamma^2-1}}.
\end{equation}
The transformation between the Weyl coordinates $(\rho,z)$ of Eq. (\ref{Weyl}) 
and the Erez-Rosen coordinates $(r,\theta)$ of Eq. (\ref{eq:metric_gamma}) is \cite{Stephani,Herrera_1999}
\begin{equation}
\rho^2=\left(r^2-2mr\right)\sin^2\theta~,~~z=\left(r-m\right)\cos\theta.
\end{equation}

The metric is characterized by two parameters $m$ and $\gamma$, where $m$ is related to the mass, 
and $ \gamma $ measures deformation of the spacetime from spherical symmetry. The 
Arnowitt-Deser-Misner (ADM) mass of the spacetime is $ M=m\gamma $, and the corresponding 
quadrupole moment is $Q=\frac{\gamma M^3}{3}(1-\gamma^2)$ \cite{Herrera_1999}. The monopole $M$ 
and the quadrupole $Q$ are the only independent components of multipole moments, as all higher 
order components can be expressed in terms of $M$ and $Q$. When $ \gamma=0 $, the $\gamma$-metric 
represents the flat Minkowski spacetime, and for $ \gamma=1 $, it reduces to the spherically symmetric 
Schwarzschild solution. For all other values of $\gamma$, it deforms the spacetime from spherically symmetric 
to axially symmetric, with $\gamma<1$ $(\gamma>1)$ representing a prolate (oblate) spheroid. 

The $\gamma$-metric is an interesting model to understand the directional behaviour of naked
singularities \cite{Herrera_2005,Virbhadra_1996}. 
Since this is a vacuum solution, the Ricci scalar for the metric vanishes. Whereas the expression of the 
Kretschmann scalar, $\mathcal{K}=R_{\alpha\beta\delta\lambda}R^{\alpha\beta\delta\lambda}$ 
($R_{\alpha\beta\delta\lambda}$ is the Riemann curvature tensor), 
reads \cite{Virbhadra_1996,Quevedo_2011,Boshkayev_2016}
\begin{equation}
\mathcal{K}=\frac{16m^2\gamma^2}{r^{2(\gamma^2+\gamma+1)}(r-2m)^
{2(\gamma^2-\gamma+1)}(r^2-2mr+m^2\sin^2\theta)^{3-2\gamma^2}}~F(r,\theta),
\label{eq:Kret_scalar}
\end{equation}
where
\begin{eqnarray}
F(r,\theta) &=& m^2\sin^2\theta\left[3m\gamma\left(\gamma^2+
1\right)(m-r)+\gamma^2\left(4m^2-6mr+3r^2\right)+m^2\left(\gamma^4+1\right)\right]\nonumber\\
&+& 3r(\gamma m+m-r)^2(r-2m).
\end{eqnarray}
It can be seen that the Kretschmann scalar diverges at $r=0$ for all $\gamma>0$. So, there is a curvature
singularity at $r=0$ for all $\gamma>0$. Moreover,  
the nature of the surface $r=2m=\frac{2M}{\gamma}$ is quite interesting. As for the Schwarzschild case, i.e., for $\gamma=1$,
this surface marks the location of the event horizon, which also represents the infinitely red-shifted surface
for observers at spatial infinity. From Eq. (\ref{eq:Kret_scalar}), the expressions of the Kretschmann scalar along the 
polar axis ($ \theta=0,\pi $) and $ \theta=\pi/2 $ become \cite{Virbhadra_1996}
\begin{equation}
\mathcal{K}\rvert_{\theta=0}=\mathcal{K}\rvert_{\theta=\pi}=\frac{48m^2\gamma^2(\gamma m+m-r)^2}{r^{4+2\gamma}(r-2m)^{4-2\gamma}},
\label{eq:Kret_pol}
\end{equation}
\begin{equation}
\mathcal{K}\rvert_{\theta=\frac{\pi}{2}}=\frac{16m^2\gamma^2}{r^{2\gamma^2+2\gamma+2}(r-2m)^{2\gamma^2-2\gamma+2}(r-m)^{6-4\gamma^2}}~H(r),
\label{eq:Krte_eq}
\end{equation}
where
\begin{align}
\nonumber H(r) ~ &= ~ m^4\left(\gamma^4+3\gamma^3+4\gamma^2+3\gamma+1\right)
-3m^3r\left(\gamma^3+4\gamma^2+5\gamma+2\right)\\
&+ ~ 3m^2r^2\left(2\gamma^2+6\gamma+5\right)-6mr^3(\gamma+2)+3r^4.
\end{align}

From Eqs. (\ref{eq:Kret_scalar}), (\ref{eq:Kret_pol}) and (\ref{eq:Krte_eq}), we see that, for $\gamma<2~(\neq 0,1)$, the 
Kretschmann scalar diverges at $r=2m=\frac{2M}{\gamma}$ for the whole domain of $\theta$, i.e., 
for $0\le\theta\le\pi$. So there is a curvature singularity at $r_s=2m=\frac{2M}{\gamma}$ for the mentioned 
range of $\gamma$. On the other hand, for $\gamma>2$ case, the Kretschmann scalar becomes 
zero at $r=2m$ at $\theta=0,\pi$ and diverges at $\theta=\pi/2$. Therefore, no singularity exists on 
the polar axis in the range $\gamma>2$, whereas along the equatorial direction, the curvature 
singularity does exist at $r=r_s$ for all values of $\gamma>0~(\neq 1)$ \cite{Galtsov_2019}. 
Moreover, the surface $r=r_s$ also represents an infinitely red-shifted surface for any value of 
$\gamma>0$ ($\gamma=1$ being the Schwarzschild one) for an observer at spatial infinity. 
For example, if $\nu$ is the frequency of a light ray emitted from a source at rest at a finite radial 
distance $r$, and $\nu_{\infty}$ is the corresponding frequency of the same light ray received by an 
observer at rest at infinity, then we can write
\begin{equation*}
\nu_{\infty}=\nu\sqrt{A(r)}=\nu\left(1-\frac{2m}{r}\right)^{\gamma/2}.
\end{equation*}
As $r \to r_s=2m$, we have $\nu_{\infty} \to 0$, i.e., the light received by the observer is infinitely red-shifted, 
irrespective of the value of $ \gamma>0 $. Again from Eq. (\ref{eq:Kret_scalar}), we observe that 
$\mathcal{K}\to\infty$ when $r^2-2mr+m^2\sin^2\theta=0$ for the range $0<\gamma\le\sqrt{3/2}$. 
Therefore, in addition to the singularities at $r=0$ and $r_s=2m$, there exists another singular 
surface inside the $r_s=2m$ singularity, for this range of $\gamma$. Since we are interested 
only in the region exterior to the outermost singular surface $r_s=2m$, the singularities internal to this 
surface are immaterial to our analysis. For associated literature regarding the properties of this spacetime, 
for example the global structure, motion of test particles, accretion disk properties etc., we refer the reader to  \cite{Hernandez_1994,Herrera_1999,Kodama_2003,Quevedo_2011,Chowdhury_2012,Boshkayev_2016}.

\subsection{$\gamma\to\infty$ limiting case}
\label{limiting}

We have also considered the spacetime in the limiting case of $ \gamma \to \infty $, keeping the ADM 
mass $ M $ fixed and finite. The resulting spacetime corresponds to the Chazy-Curzon solution of GR 
in spherical coordinates \cite{Griffiths_2009,Chazy_1924,Curzon_1924,Camacho_2015}. We call it as 
Gamma-Infinite (GI) spacetime for simplicity. The corresponding line element of the GI spacetime is
\begin{equation}
ds^2 = -\exp\left(-\frac{2M}{r}\right)dt^2 + \exp\left(\frac{2M}{r}\right)\left[\exp\left(-\frac{M^2\sin^2\theta}{r^2}\right)
\left(dr^2 + r^2 d\theta^2\right) + r^2\sin^2\theta d\phi^2\right].
\label{eq:metric_GI}
\end{equation}
The Kretschmann scalar of this GI spacetime reads
\begin{equation}
\tilde{\mathcal{K}}=\frac{16M^2\left[3(M-r)^2r^2+M^2(M^2-3Mr+3r^2)\sin^2\theta\right]}{r^{10}}\exp
\left[\frac{2M(M\sin^2\theta-2r)}{r^2}\right].
\end{equation}
From the above expression, we see that
\begin{equation}
\left.\tilde{\mathcal{K}}\right|_{r\to 0}=\left\{ \begin{array}{ll}
0, & \mbox{if $~~\theta=0,\pi$}\\
\infty, & \mbox{if $~~\theta \neq 0,\pi$}\end{array} \right.
\end{equation}
Therefore, there is no curvature singularity at $ r=0 $ for $ \theta=0,\pi $, whereas for all other values of 
$ \theta $, curvature singularity exists at $ r=0 $. We shall also discuss the lensing properties of this spacetime in sequel.

\section{Optical properties and shadows of the $\gamma$-metric}
\label{shadow}

To analyze the gravitational lensing and  shadows cast by the $\gamma$-metric, we need to study the motion of 
photons in this background. Due to time translational and azimuthal symmetries of this spacetime given by the 
metric in Eq. (\ref{eq:metric_gamma}), we have two constants of motion along the null geodesics, namely, the 
energy $E$ and the angular momentum $L$ of photons about the axis of symmetry. Therefore, the $t$ and $\phi$ 
components of the geodesic equations for photons are given by
\begin{equation}
\dot{t}=-\frac{E}{g_{tt}}, \quad \dot{\phi}=\frac{L}{g_{\phi\phi}}.
\end{equation}
where an `overdot' represents differentiation with respect to the affine parameter (unless otherwise specified). 
From the normalization of four velocities of photons ($u^{\mu}u_{\mu}=0$) along null geodesics, we obtain
\begin{equation}
(-g_{tt})\left(g_{rr}\dot{r}^2+g_{\theta\theta}\dot{\theta}^2\right)+\left(\frac{-g_{tt}}{g_{\phi\phi}}\right)L^2=E^2~,~
\text{i.e.,}~ (-g_{tt})\left(g_{rr}\dot{r}^2+
g_{\theta\theta}\dot{\theta}^2\right)+
V_{\text{eff}}=E^2
\label{eq:normalization_photon}
\end{equation}
where $V_{\text{eff}}$ is the effective potential for photons and is given by
\begin{equation}
V_{\text{eff}}=\frac{L^2}{r^2\sin^2\theta}\left(1-\frac{2m}{r}\right)^{2\gamma-1}=\frac{L^2}{r^2\sin^2\theta}
\left(1-\frac{2M}{\gamma r}\right)^{2\gamma-1}.
\label{eq:eff_pot_photon}
\end{equation}

\begin{figure}[h]
\includegraphics[scale=0.9]{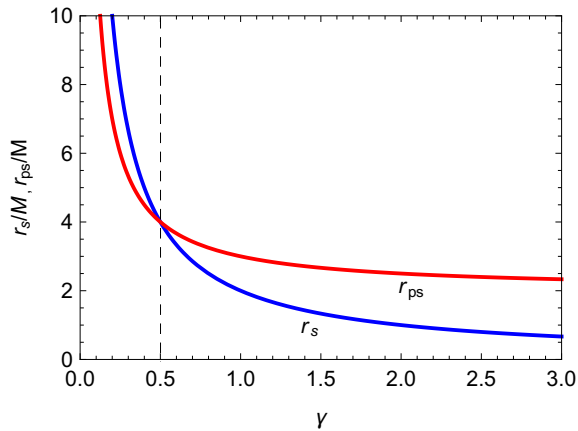}
\caption{Variation of the photon capture radius $(r_{\text{ps}})$ and the singularity $(r_s)$ on the 
equatorial plane as functions of $\gamma$. The solid red line corresponds to $r_{\text{ps}}$ and the 
solid blue line represents $r_s$. We have chosen $M=1$ to obtain the plots. For $\gamma<0.5$, 
we see that $r_{\text{ps}}<r_s$, i.e., the singular surface is outside the photon capture circle.}
\label{fig:Photon_sph}
\end{figure}

\begin{figure}[h]
\includegraphics[scale=0.75]{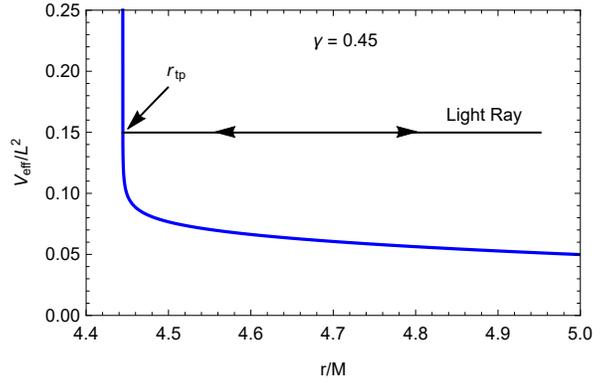}
\caption{A typical plot of $V_{\text{eff}}^{\pi/2}$ (in units of $\bar{L}^2$) as a function of $r$ (in units of $M$) 
for $\gamma=0.45$. A ray of light (indicated by the black line) comes from a distant source, hits the turning 
point at $r=r_{\text{tp}}$, turns back and escapes to infinity.}
\label{fig:Effective_pot}
\end{figure}

On the equatorial plane, $\theta=\pi/2$ and $\dot{\theta}=0$. Therefore, on this plane, we have
\begin{equation}
(-g_{tt})~g_{rr}\dot{r}^2+V_{\text{eff}}^{\pi/2}=E^2
\label{eq:rdot_photon_equatorial}
\end{equation}
where the effective potential on the equatorial plane takes the form
\begin{equation}
V_{\text{eff}}^{\pi/2}=\frac{L^2}{r^2}\left(1-\frac{2m}{r}\right)^{2\gamma-1}=\frac{L^2}{r^2}\left(1-\frac{2M}{\gamma r}\right)^{2\gamma-1}.
\label{eq:eff_pot_photon_equtorial}
\end{equation}
If a photon arrives at the turning point ($r_{\text{tp}}$) of its trajectory, we have $\dot{r}\rvert_{r=r_{\text{tp}}}=0$, 
which from Eqs. (\ref{eq:rdot_photon_equatorial}) and (\ref{eq:eff_pot_photon_equtorial}) gives, 
\begin{equation}
\left.V_{\text{eff}}^{\pi/2}~\right\rvert_{r=r_{\text{tp}}}=\frac{L^2}{r_{\text{tp}}^2}\left(1-\frac{2M}
{\gamma r_{\text{tp}}}\right)^{2\gamma-1}=E^2 ~~ \implies ~~ b(r_{\text{tp}})=\frac{L}{E}=
r_{\text{tp}}\left(1-\frac{2M}{\gamma r_{\text{tp}}}\right)^{(1-2\gamma)/2}
\label{eq:impact_parameter}
\end{equation}
where $ b(r_{\text{tp}})$ is the impact parameter of a light ray. The maximum of the effective potential 
marks the position of the photon capture radius which is known as the light ring in an axially symmetric case, 
and the photon sphere in a spherically symmetric case. On the equatorial plane, the photon capture radius becomes
\begin{equation}
r_{\text{ps}}=(2\gamma+1)m=2M+\frac{M}{\gamma}.
\label{eq:ps_gamma}
\end{equation}
Figure \ref{fig:Photon_sph} shows the variation of the photon capture radius $r_{\text{ps}}$ on the 
equatorial plane and the singularity radius $r_s$  as functions of $\gamma$. Note that the photon capture 
radius exists only for $\gamma\geq 1/2$. However, for $\gamma<1/2$, it does not exist as $r_{\text{ps}}<r_s$ in this case. 
This is also clear from the effective potential $V_{\text{eff}}^{\pi/2}$. For $\gamma>1/2$, $V_{\text{eff}}^{\pi/2}$ 
vanishes both at the singularity $r_s$ and at the spatial infinity with a maximum in between marking the 
position of $r_{\text{ps}}$. However, for $\gamma<1/2$, $V_{\text{eff}}^{\pi/2}$ diverges at the singularity 
$r_s$ (see Fig. \ref{fig:Effective_pot}). Therefore, in this latter case, photons on the equatorial plane will 
always have turning points outside the singularity. This is also true for off-equatorial photon geodesics. 
The reason is that, if the geodesics governed by Eq. (\ref{eq:rdot_photon_equatorial}) always have 
turning points, then so do the ones governed by Eq. (\ref{eq:normalization_photon}), as, in the limit $r\to r_s$, the effective potential in Eq. (\ref{eq:normalization_photon}) still diverges and the coefficient of $\dot{\theta}^2$ vanishes. Therefore, in the $\gamma<1/2$ case, as a photon with any non-zero impact parameter always has a turning point, there will be no capturing of photons at all, and hence no shadow will be produced in this case.

To check the above results, we use our numerical ray-tracing techniques (with certain modifications) discussed in 
our previous work \cite{Suvankar_2020} and produce the images. We shoot photons with 
different impact parameters from a distant observer towards the lensing objects and integrate the geodesic equations 
backward in time. If a geodesic has a turning point and escapes to infinity for a given impact parameter, 
we assign bright point to this. On the other hand, if it is captured by the singularity, we assign dark point to the 
corresponding impact parameter. However, since at the singularity $r_s=\frac{2M}{\gamma}$ some of the metric 
functions become infinite, we cannot exactly touch the surface of singularity due to numerical limitation. We 
shall have to take a region of tolerance $(\delta r)$ around the singularity. Therefore, we take the inner grid to be 
at $r=r_s+\delta r$, where $\delta r$ is very small. Any photon hitting this surface is assumed to be captured 
by the singularity. Also, while performing ray-tracing, we consider piecewise step size in the affine parameter 
$\lambda$. When the radial coordinate reaches below a predefined value $r_s+0.3$, i.e., when $r\leq (r_s+0.3)$ 
along a geodesic, we decrease the step size to $\Delta\lambda=10^{-5}$.

\begin{figure}
\centering
\subfigure[~$\gamma\to \infty$ (GI)]{\includegraphics[scale=0.47]{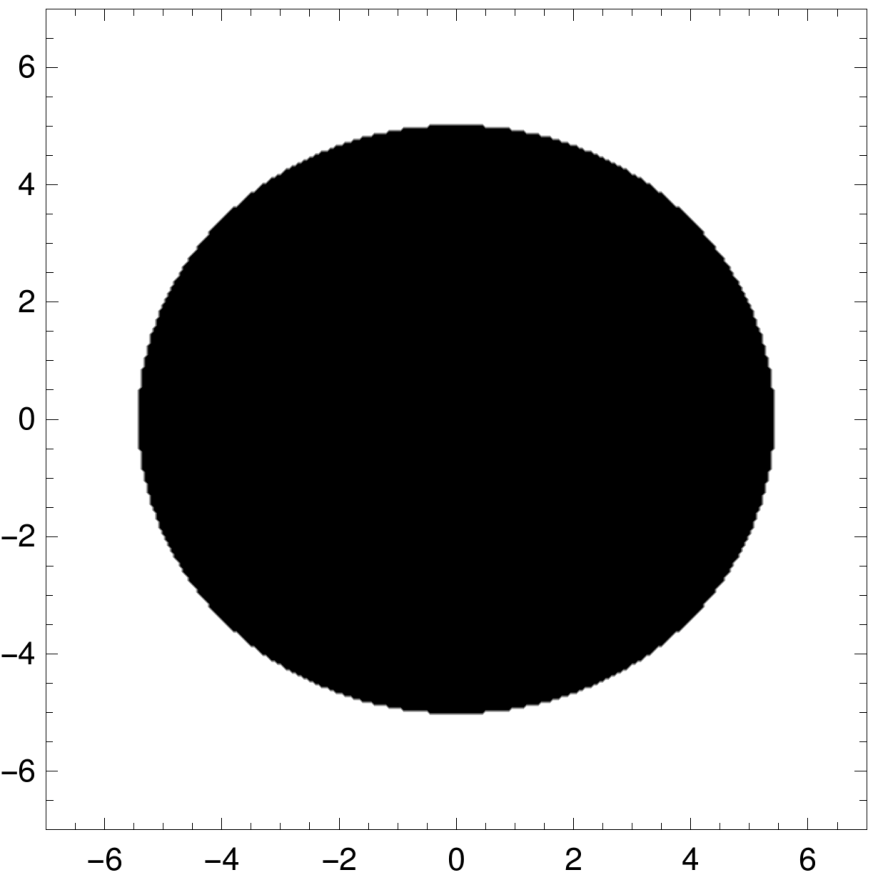}}
\subfigure[~$\gamma=1.0$ (Schwarzschild)]{\includegraphics[scale=0.47]{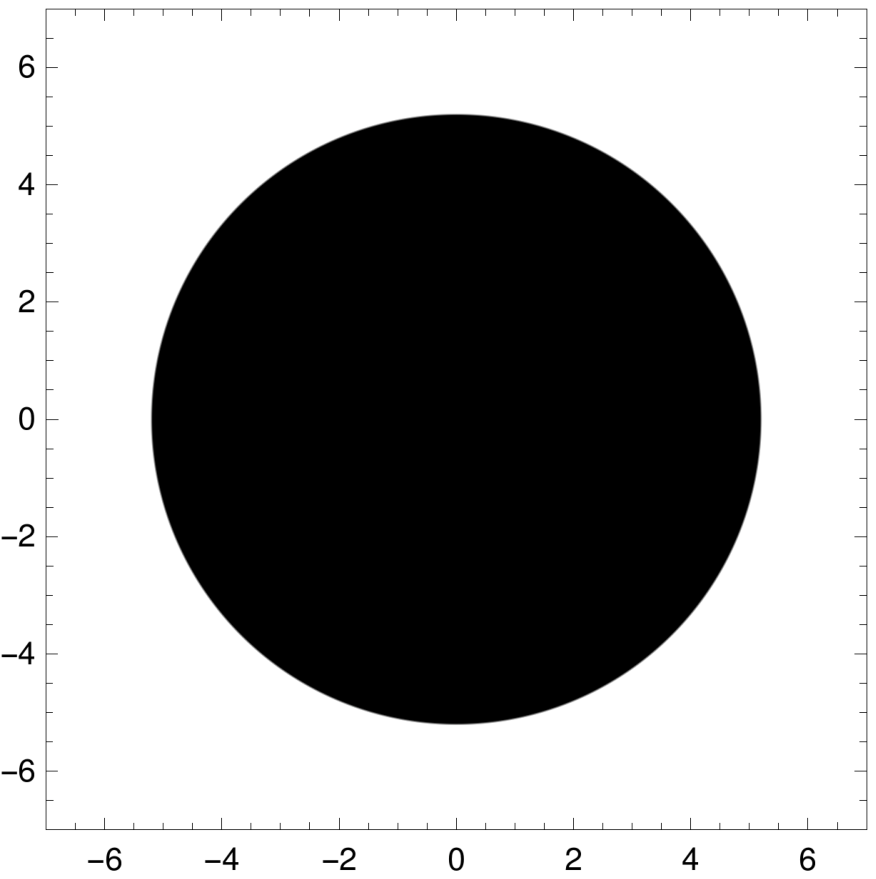}}
\subfigure[~$\gamma=0.75$]{\includegraphics[scale=0.47]{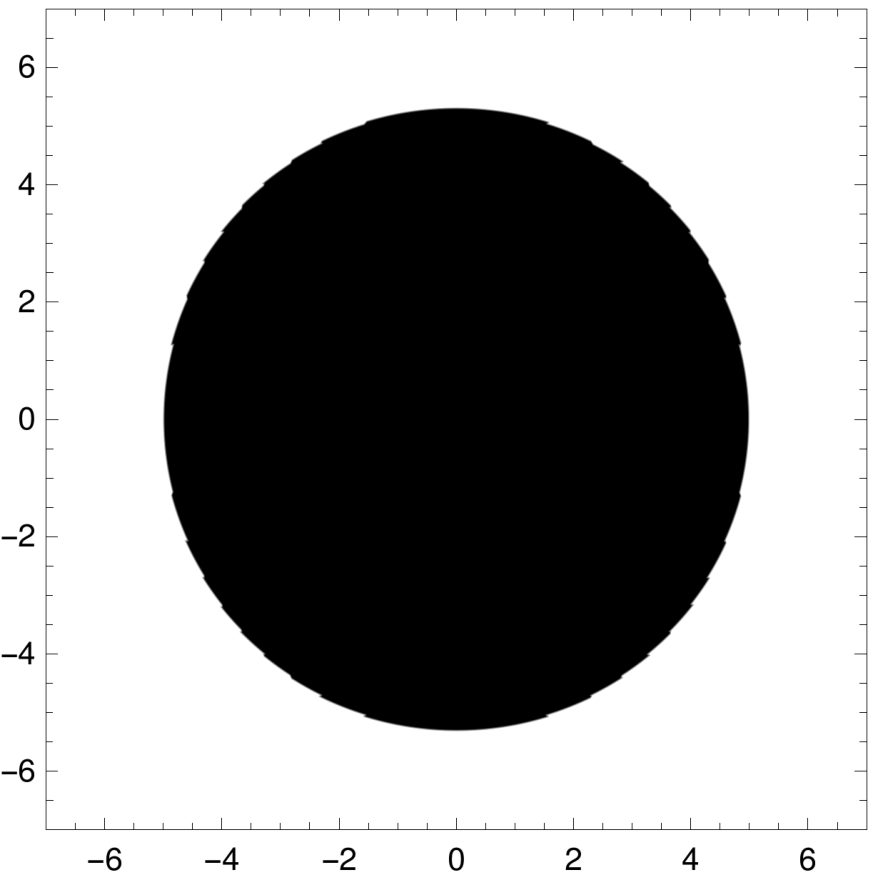}}
\subfigure[~$\gamma=0.51$]{\includegraphics[scale=0.47]{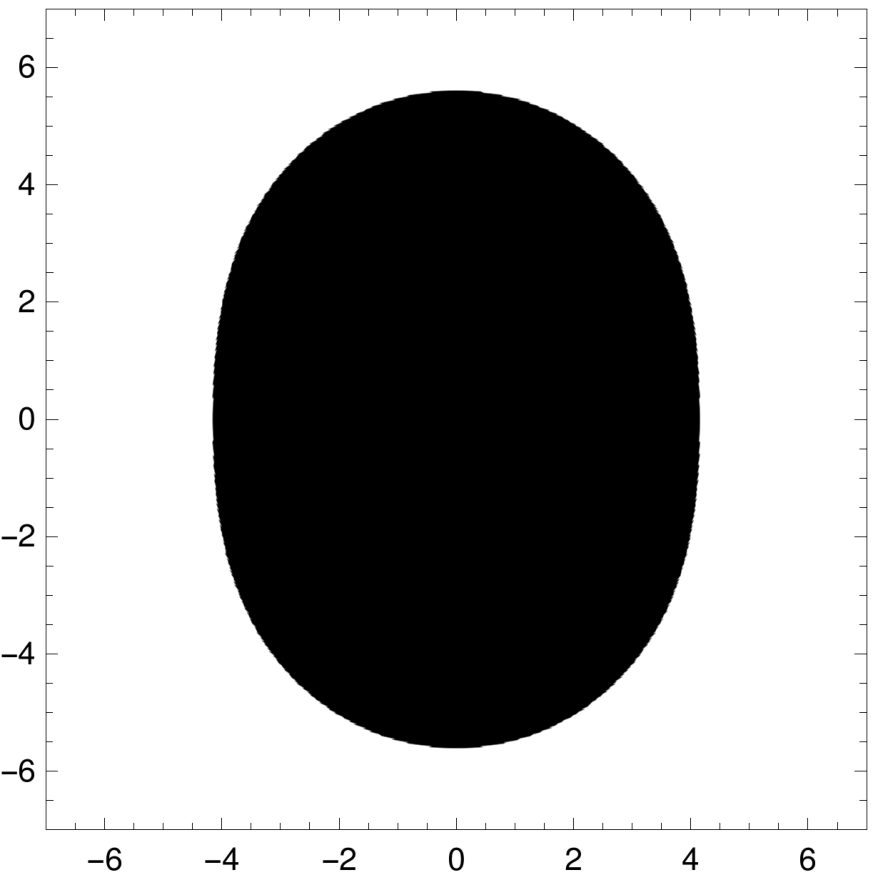}}
\subfigure[~$\gamma=0.4$, $\delta r=0.05$]{\includegraphics[scale=0.47]{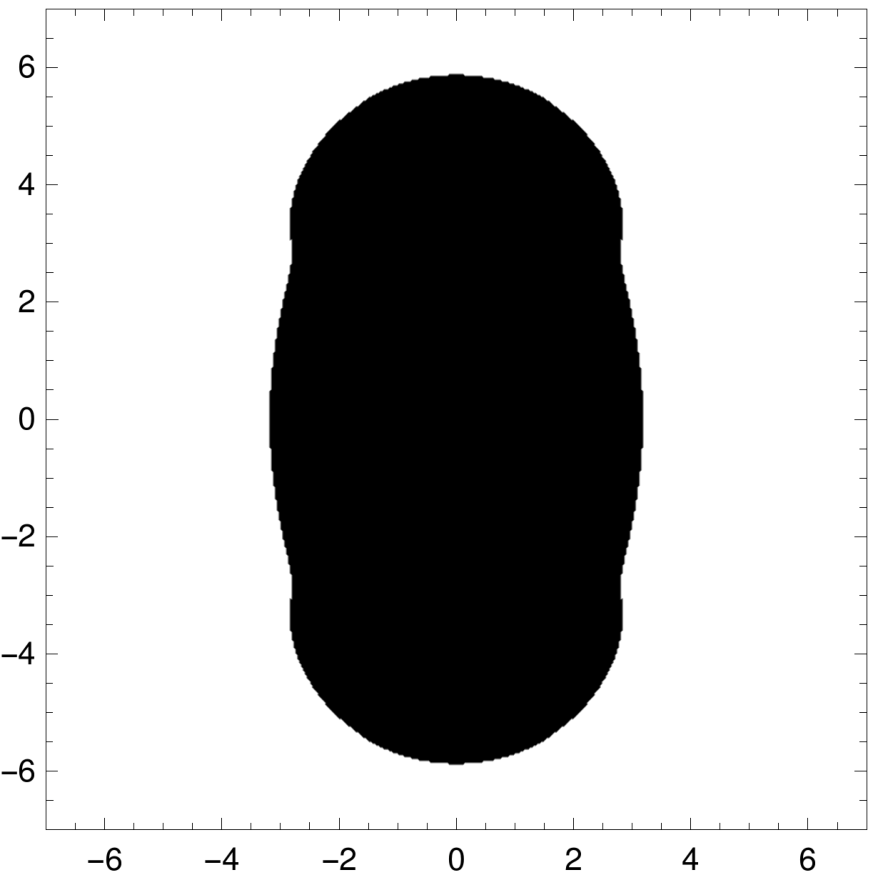}}
\subfigure[~$\gamma=0.4$, $\delta r=0.001$]{\includegraphics[scale=0.47]{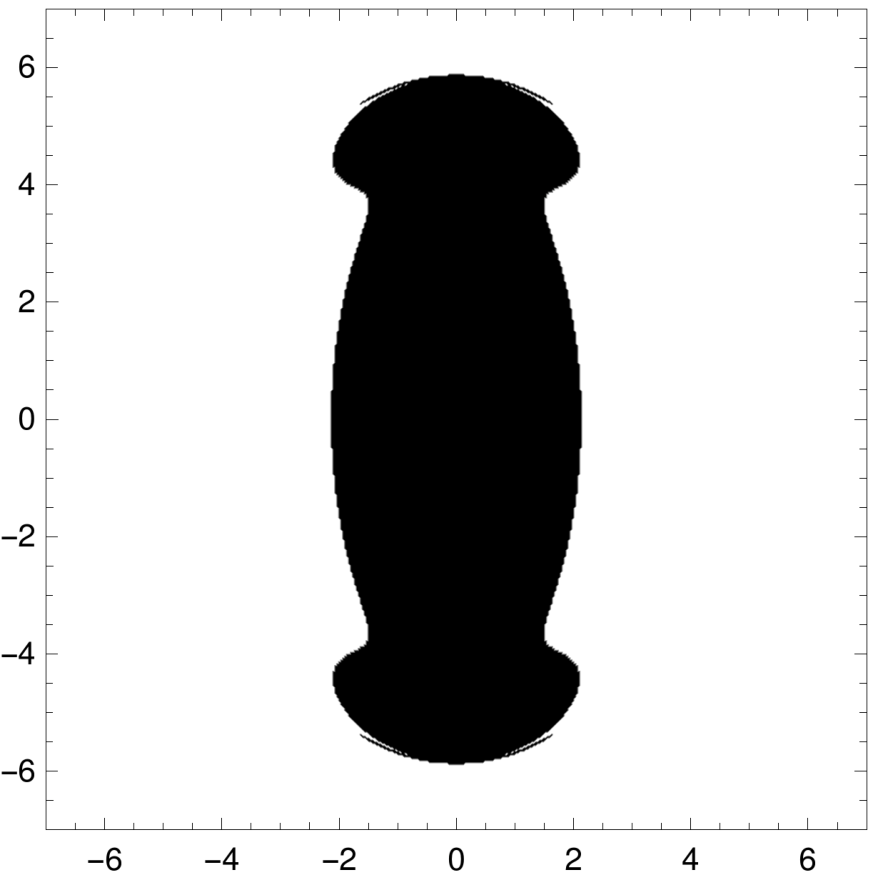}}
\subfigure[~$\gamma=0.4$, $\delta r=10^{-6}$]{\includegraphics[scale=0.47]{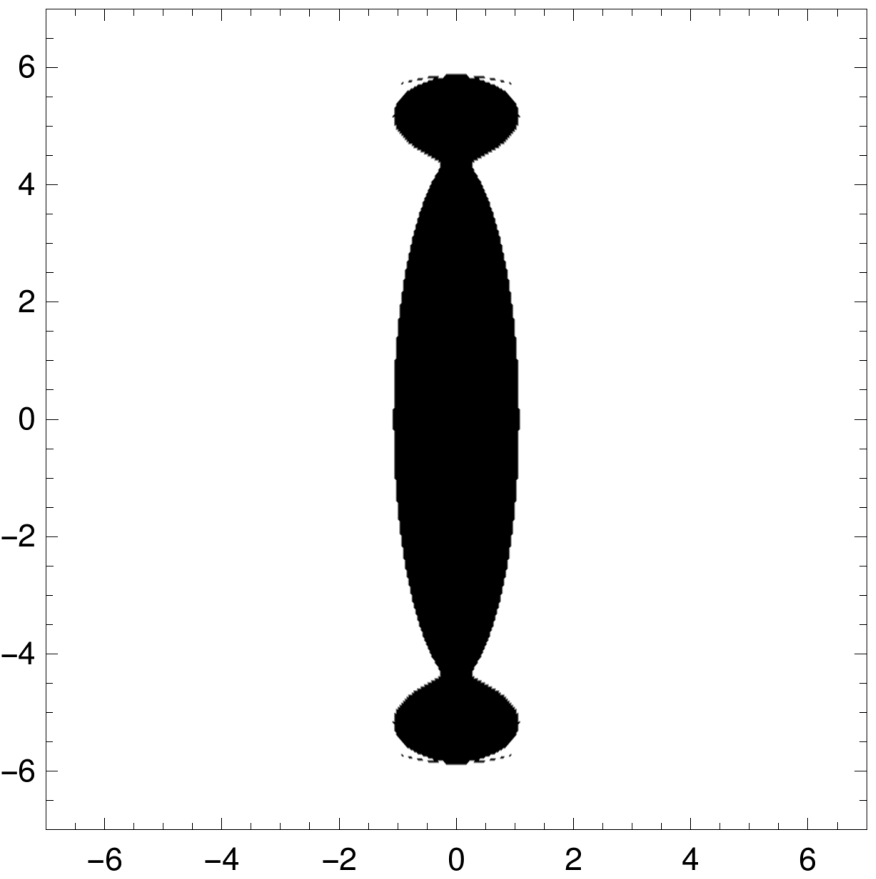}}
\subfigure[~$\gamma=0.4$, $\delta r=10^{-9}$]{\includegraphics[scale=0.47]{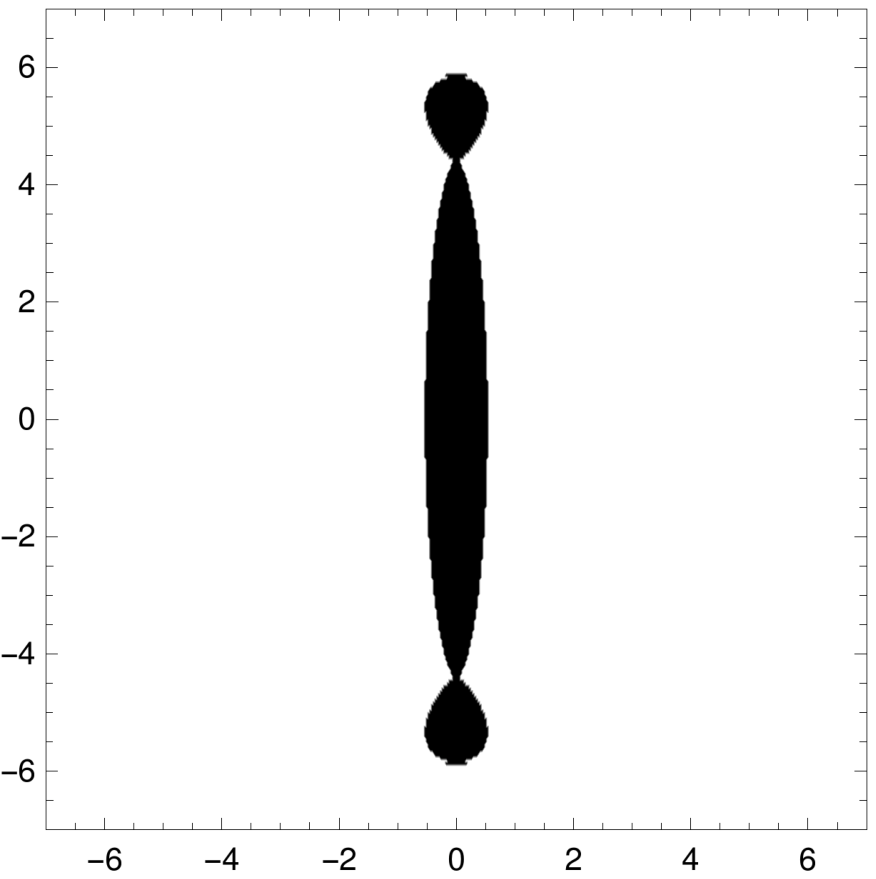}}
\caption{Ray-traced shadows of the $\gamma$-metric. The black region will disappear for $\gamma<1/2$ 
in the limit $\delta r\to 0$, implying that there is no shadow in this case. The observer's inclination angle is taken to be $\theta_o=\pi/2$.}
\label{fig:Shadows_gamma}
\end{figure}

Figure \ref{fig:Shadows_gamma} shows the ray-traced shadows of the $\gamma$-metric for different $\gamma$. 
Note that, for $\gamma\geq 1/2$, the singularity always casts shadows whose shapes and sizes will depend on 
the value of $\gamma$. The shadow becomes more and more prolate as we decrease $\gamma$. This is similar to 
the results obtained in \cite{Bambi_2019,Galtsov_2019}. However, for $\gamma=0.4$, we note that the dark region 
shrinks as we decrease $\delta r$ and will vanish in principle in the limit $\delta r\to 0$. Therefore, as explained before
via a theoretical argument, $\gamma<1/2$ case does not cast any shadow. This conclusion is in contradiction to the 
studies of shadows in \cite{Bambi_2019,Galtsov_2019}.

\begin{figure}
\centering
\subfigure{\includegraphics[scale=0.8]{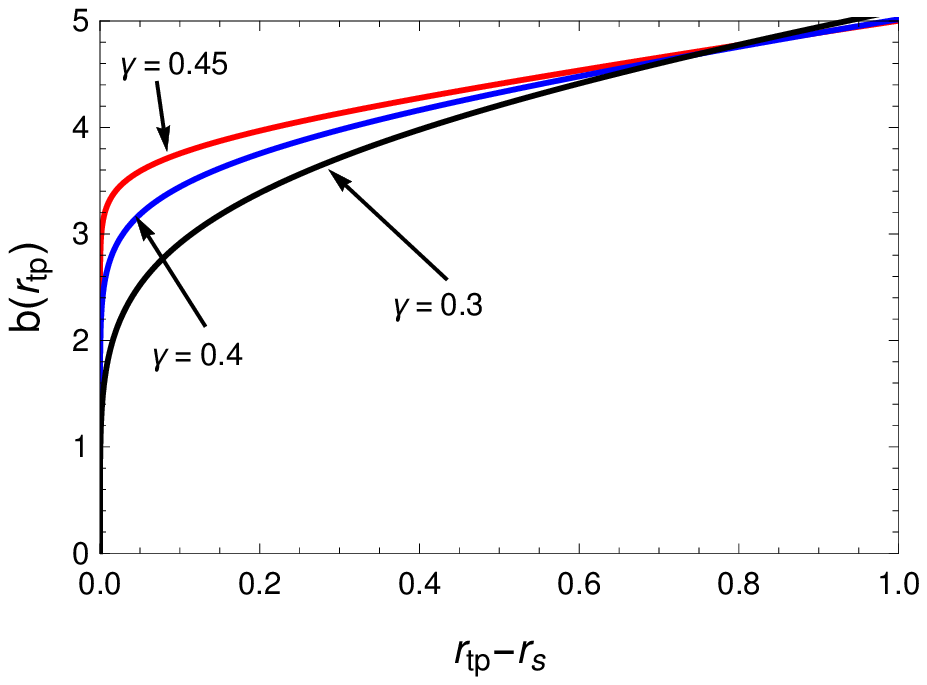}}
\subfigure{\includegraphics[scale=0.8]{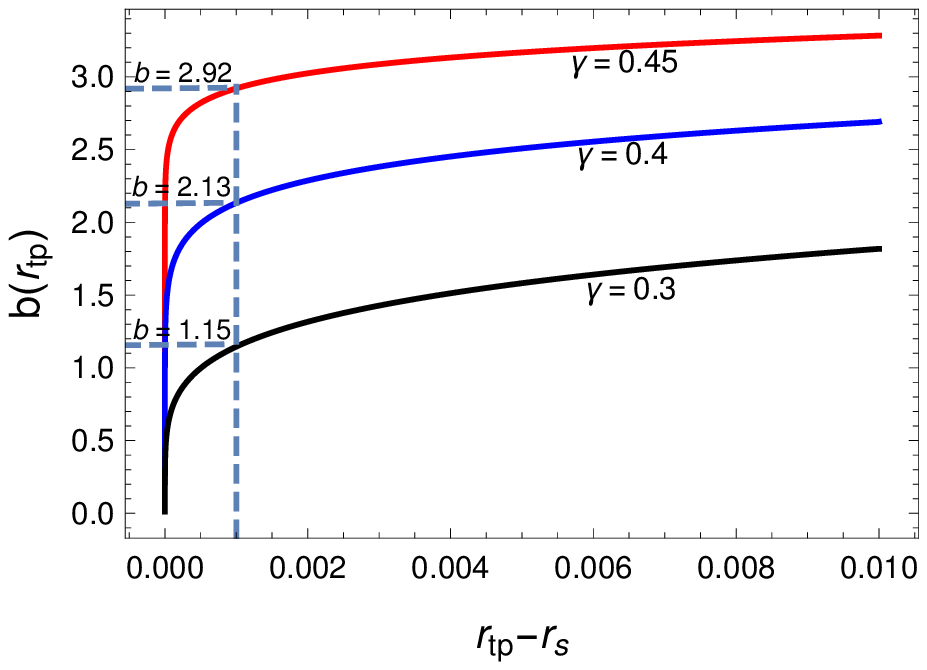}}
\caption{Impact parameter as a function of the turning point $r_{tp}$ on the equatorial plane. The 
vertical dashed line corresponds to turning point $r_{tp}=r_s+0.001$. The corresponding impact 
parameters for this turning point are $2.92$ $2.13$ and $1.15$ for $\gamma=0.45$, $0.4$ and $0.3$, respectively. Here, we have considered $M=1$.}
\label{fig:btp}
\end{figure}

The shrinking of the dark region with decreasing $\delta r$ can be understood in more detail 
from the analysis of geodesics on the equatorial plane. In Fig. \ref{fig:btp}, we have shown dependence of 
the impact parameter on turning point $r_{tp}$ for equatorial geodesics. Note that the turning point lies 
very close to the singularity even when the impact parameter is $\sim$ $3$ or $4$. If we take the turning point 
to be $r_{tp}=r_s+0.001$, then corresponding impact parameters are $2.92$ $2.13$ and $1.15$ for 
$\gamma=0.45$, $0.4$ and $0.3$, respectively. Therefore, if we take $\delta r=0.001$, then this means that 
we are excluding all those photons having turning points $r_{tp}\leq (r_s+0.001)$, i.e., from the impact 
parameters space, we are excluding impact parameters $b\leq 2.92$, $2.13$ and $1.15$ for 
$\gamma=0.45$, $0.4$ and $0.3$, respectively. Photons having impact parameters within the excluded 
region form dark spots. As a result, we are having the dark region for $\gamma<1/2$. However, for a 
given $\gamma<1/2$, decreasing $\delta r$ means decreasing the excluded region from the impact 
parameter space. The excluded region and hence the dark region vanishes in the limit $\delta r\to 0$. 
Therefore, we do not have any shadow in principle in the $\gamma<1/2$ case.

\section{Constraining the $\gamma$-metric using the M87$^*$ results}
\label{constrain}

We now use the results from M87$^*$ observation \citep{EHT1} and put possible constraint on the $\gamma$-metric. 
For this purpose, we use the average size of the shadow and its deformation from circularity. To this end, 
we first denote the horizontal and the vertical axes in the shadow plane by $\alpha$ and $\beta$ respectively 
and define an angle $\phi$ between the $\alpha$-axis and the vector connecting the geometric centre $(0,0)$ 
of the shadow with a point $(\alpha,\beta)$ on the boundary of a shadow. Therefore, the average radius 
$R_{av}$ of the shadow is given by \citep{Bambi}
\begin{equation}
R_{av}^2=\frac{1}{2\pi}\int_{0}^{2\pi}l^2(\phi)\; d\phi,
\end{equation}
where $l(\phi)=\sqrt{\alpha(\phi)^2+\beta(\phi)^2}$ and $\phi=tan^{-1}(\beta(\phi)/\alpha(\phi))$. 
Following \cite{EHT1}, we define the deviation $\Delta C$ from circularity as
\begin{equation}
\Delta C=\frac{1}{R_{av}}\sqrt{\frac{1}{2\pi}\int_{0}^{2\pi}(l(\phi)-R_{av})^2\; d\phi}.
\end{equation}
Note that $\Delta C$ is the fractional RMS distance from the average radius of the shadow.
\begin{figure}[h!]
\centering
\subfigure{\includegraphics[scale=0.8]{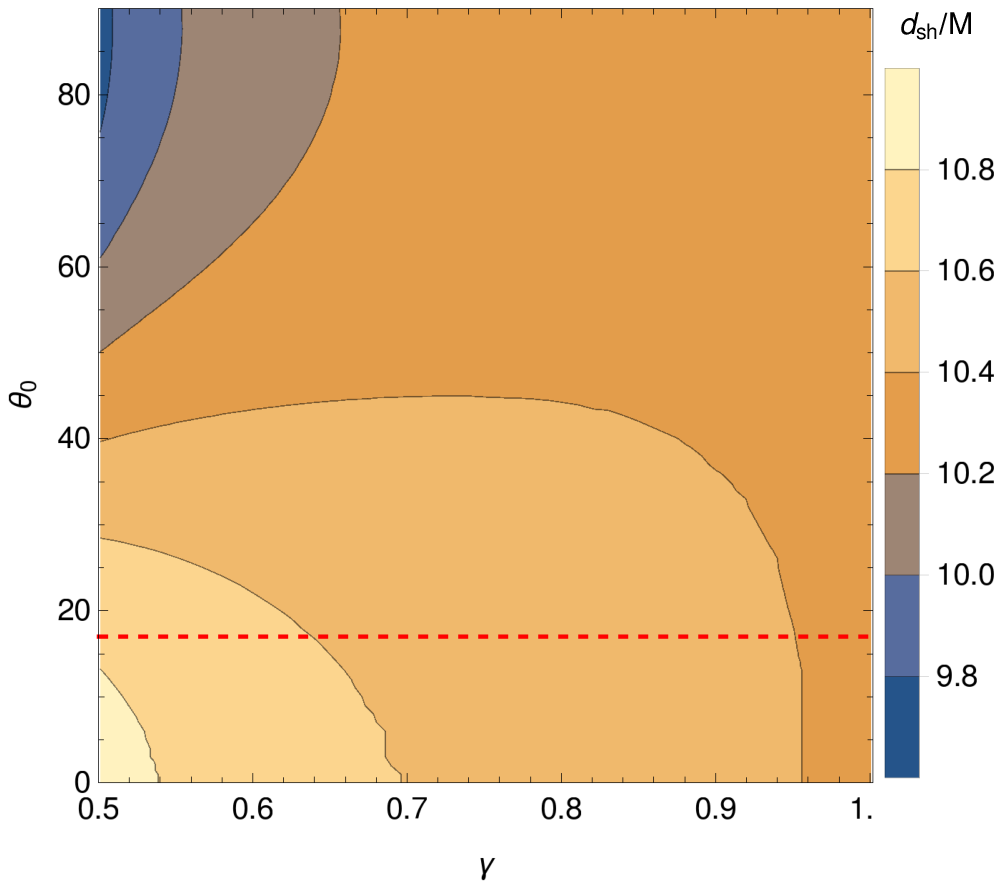}}
\subfigure{\includegraphics[scale=0.8]{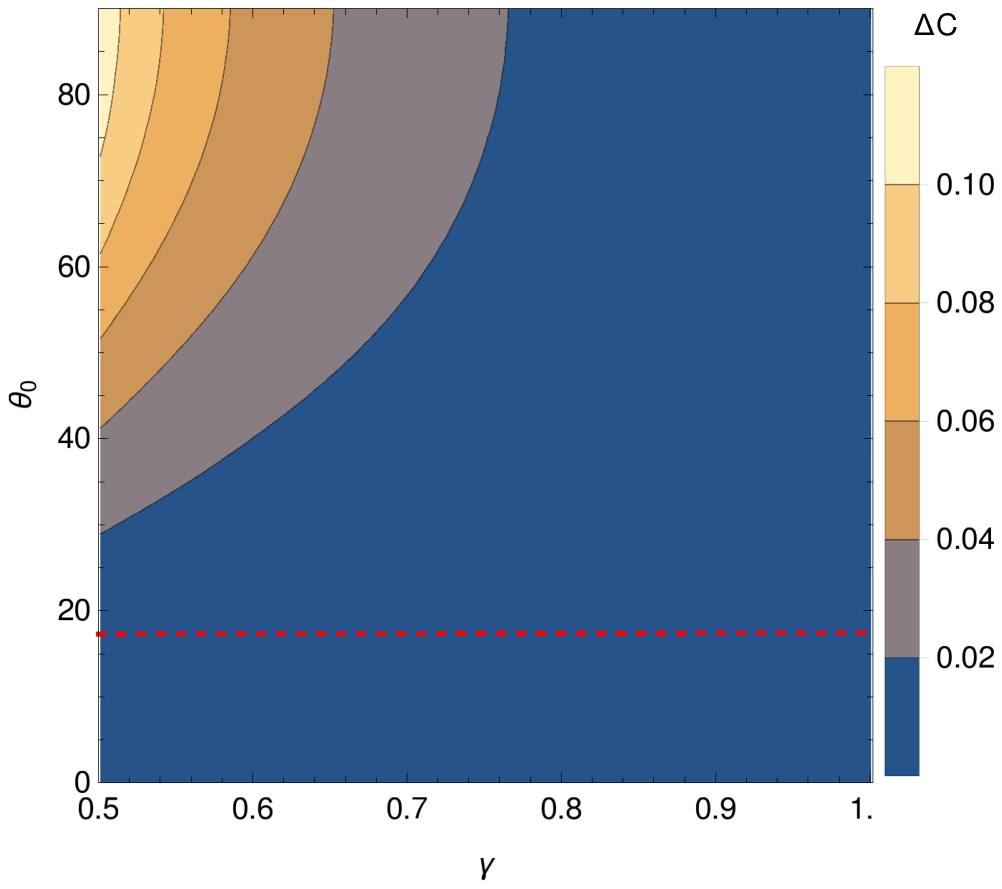}}
\caption{Dependence of the diameter and deformation of the shadow on $\gamma$ and $\theta_o$. Here, $M=(6.5\pm 0.7)\times 10^9 M_\odot$ and $D=(16.8\pm 0.8)$ Mpc. The red dashed line shows the inclination angle $\theta_o=17^\circ$. Note that, for this inclination angle, the diameter and the deformation of the shadow are consistent with the 
M87$^*$ observation.}
\label{fig:constraint}
\end{figure}
According to EHT collaboration, the angular size of the observed shadow is $\Delta\theta_{sh}=42\pm 3$ $\mu$as, 
and its deviation from circularity ($\Delta C$) is less than $10\%$ \citep{EHT1}. Also, following \cite{EHT1}, we take 
the distance to M87$^*$ to be $D=(16.8\pm 0.8)$ Mpc and the mass of the object to be 
$M=(6.5\pm 0.7)\times 10^9 M_\odot$. These numbers imply that the diameter of the shadow in dimensionless unit should be
\begin{equation}
\frac{d_{sh}}{M}=\frac{D\Delta\theta_{sh}}{M}=11.0\pm 1.5,
\end{equation}
where the errors have been added in quadrature. The above quantity must be equal to $\frac{2R_{av}}{M}$, i.e., $\frac{d_{sh}}{M}=\frac{2R_{av}}{M}$. 
In Fig. \ref{fig:constraint}, we have shown the average diameter and the deviation from circularity of the shadow 
for different $\gamma$ and the observers inclination angle $\theta_o$. Here, we have taken $0.5\leq \gamma\leq 1$. 
Note that the size of the shadow is consistent with the M87$^*$ observations for all $\gamma$ and $\theta_o$. 
However, the deviation from circularity is more than $10\%$, i.e., $\Delta C>0.1$ over a small parameter 
region where $\gamma$ is close to $0.5$ and inclination is high simultaneously. If we restrict the inclination 
angle to be $\theta_o=17^\circ$, which the jet axis makes to the line of sight \citep{EHT1}, then both the 
shadow size and the deviation from circularity are consistent with the M87$^*$ observations for all $\gamma$ 
values considered in Fig. \ref{fig:constraint}. For the $\gamma>1$, we have found that both the size and the 
deviation from circularity slowly increases with increasing $\gamma$ for a given inclination angle. Therefore, in this 
latter case, the maximum size and deviation occur for the GI spacetime. We have found that, for 
$\theta_o=17^\circ$ and $\gamma>1$, these maximum values are given by $d_{sh}/M\simeq 10.46$ and 
$\Delta C\simeq 0.028$, which are consistent with the observation. Therefore, we find that, for the inclination 
angle of $17^\circ$, the shadow of the $\gamma$-metric is always consistent with the M87$^*$ 
observations for all $\gamma\geq 1/2$.

\section{Accretion disks and their images}
\label{accretion}

\begin{figure}[h]
\includegraphics[scale=0.75]{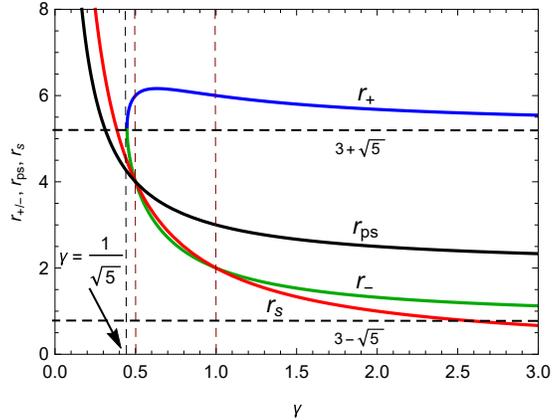}
\caption{Variations of $r_{+}$ (blue), $r_{-}$ (green), $r_s$ (red) and $r_{ps}$ (black) as functions of $\gamma$. We have considered $M=1$ to obtain the plots.}
\label{fig:ISCO}
\end{figure}

We now consider the properties and image of a  geometrically thin accretion disk in the 
$\gamma$-metric given in Eq. (\ref{eq:metric_gamma}). The disk consists of massive particles moving in stable 
circular timelike geodesics\footnote{Strictly speaking, the particles move on almost circular geodesics 
and are very slowly infalling} on the equatorial ($\theta=\pi/2$) plane and is described by the 
Novikov-Thorne model \citep{novikov_1973,page_1974}. Since the spacetime has time translational and 
azimuthal symmetries, we have two constants of motion along a timelike geodesic, namely, the specific 
energy $\tilde{E}$ (energy per unit mass) and the specific angular momentum $\tilde{L}$ about the axis 
of symmetry respectively. Therefore, the geodesic equations corresponding to $ t $ and $ \phi $ can be written as,

\begin{equation}
\dot{t}=-\frac{\tilde{E}}{g_{tt}}, \quad \dot{\phi}=\frac{\tilde{L}}{g_{\phi\phi}}.
\label{eq:t_phi_massive}
\end{equation}
From the normalization of four velocity $u^{\mu}$ (i.e. $u^{\mu}u_{\mu}=-1$) for massive particles, the radial 
geodesic equation on the equatorial plane can be written as

\begin{equation}
g_{rr}\dot{r}^2=-1-\frac{\tilde{E}^2g_{\phi\phi}+\tilde{L}^2g_{tt}}{g_{tt}
g_{\phi\phi}}=\tilde{V}_{\text{eff}}(r,\theta,\tilde{E},\tilde{L}),
\end{equation}
where $\tilde{V}_{eff}$ is the effective potential. A stable circular orbit is given by $\tilde{V}_{\text{eff}}=0$, 
$\tilde{V}'_{\text{eff}}=0$ and $\tilde{V}_{\text{eff}}''<0$. The first two conditions yield the following expressions 
for the specific energy and the specific angular
momentum of the particles moving in the stable circular orbits:

\begin{equation}
\tilde{E}=-\frac{g_{tt}}{\sqrt{-\left(g_{tt}+g_{\phi\phi}\Omega^2\right)}}, \quad\quad \tilde{L}=\frac{g_{\phi\phi}\Omega}
{\sqrt{-\left(g_{tt}+g_{\phi\phi}\Omega^2\right)}}, \quad\quad \Omega=\frac{d\phi}{dt}=\sqrt{-\frac{g_{tt,r}}{g_{\phi\phi,r}}},
\label{eq:E_L_massive}
\end{equation}
where $\Omega = d\phi/dt$ is the angular velocity of the
particles forming the disk.  The flux of the electromagnetic radiation emitted from a radial position $r$ of the disk is 
given by the standard formula \citep{novikov_1973,page_1974}
\begin{equation}
\mathcal{F}(r)=-\frac{\dot{M}}{4\pi\sqrt{-g}}\frac{\Omega'}{(\tilde{E}-\Omega \tilde{L})^2}\int_{r_{in}}^r 
(\tilde{E}-\Omega \tilde{L})\tilde{L}' dr,
\label{eq:flux}
\end{equation}
where $\dot{M}=dM/dt$ is the mass accretion rate, $r_{in}$ is the inner edge of the disk, and
$\sqrt{-g}=\sqrt{-g_{tt}g_{rr}g_{\phi\phi}}$ is the determinant of the metric on the equatorial plane. 
The marginally stable circular orbit is given by $V_{\text{eff}}''=0 $. This gives
\begin{equation}
r_{\pm}=\frac{1+3\gamma \pm \sqrt{5\gamma^2-1}}{\gamma}M.
\label{eq:ICSO_gamma}
\end{equation}
Figure \ref{fig:ISCO} shows the variations of $r_{\pm}$, the radius of singularity $r_s=\frac{2M}{\gamma}$ and the photon capture radius $r_{ps}$ as functions of $\gamma$. For $0<\gamma< 1/\sqrt{5}$, $r_{\pm}$ do not exist. Therefore, in this case, we have a single continuous disk with its inner edge at $r_{in}=r_s$ and outer edge at some radius $r>r_s$. For $\gamma=1/\sqrt{5}$, $r_{-}=r_{+}$, i.e., the outer edge of the inner disk coincides with the inner edge of the outer disk, thereby giving a single continuous disk. For $1/\sqrt{5}< \gamma < 1/2$, $r_s< r_{-}<r_{+}$, implying that there exist stable circular orbits in between the singularity $r_s$ and $r_{-}$, and also at radii greater than $r_{+}$ with no stable circular orbits in between $r_{-}$ and $r_{+}$. Therefore, in this case, we have double disk configuration (two concentric disjoint disks). The inner disk extends from the singularity (i.e., $r_{in}=r_{s}$) to $r_{-}$, and the outer disk extends from $r_{in}=r_{+}$ to some radius $r>r_{+}$. For $1/2\leq\gamma\leq 1$, $r_{-}\leq r_s<r_{+}$. Therefore, in this case, $r_{-}$ does not exist, and we have a single accretion disk with its inner edge at $r_{in}=r_{+}$ and the outer edge at some radius $r>r_{+}$. Although both the roots $r_{\pm}$ are real for $\gamma>1$ case, the angular momentum $\tilde{L}$ of circular orbits with radii $r\leq r_{-}$ becomes imaginary. Therefore, in $\gamma>1$ case, we have a single disk with its inner edge at $r_{in}=r_{+}$. 
Note that, in the limit $\gamma\to\infty$, i.e., for the GI spacetime, $r_s=0$ and $r_{\pm}=(3\pm\sqrt{5})M$. 

\begin{figure}[ht]
\centering
\subfigure[~$\gamma=1.0$ (Schwarzschild)]{\includegraphics[scale=0.54]{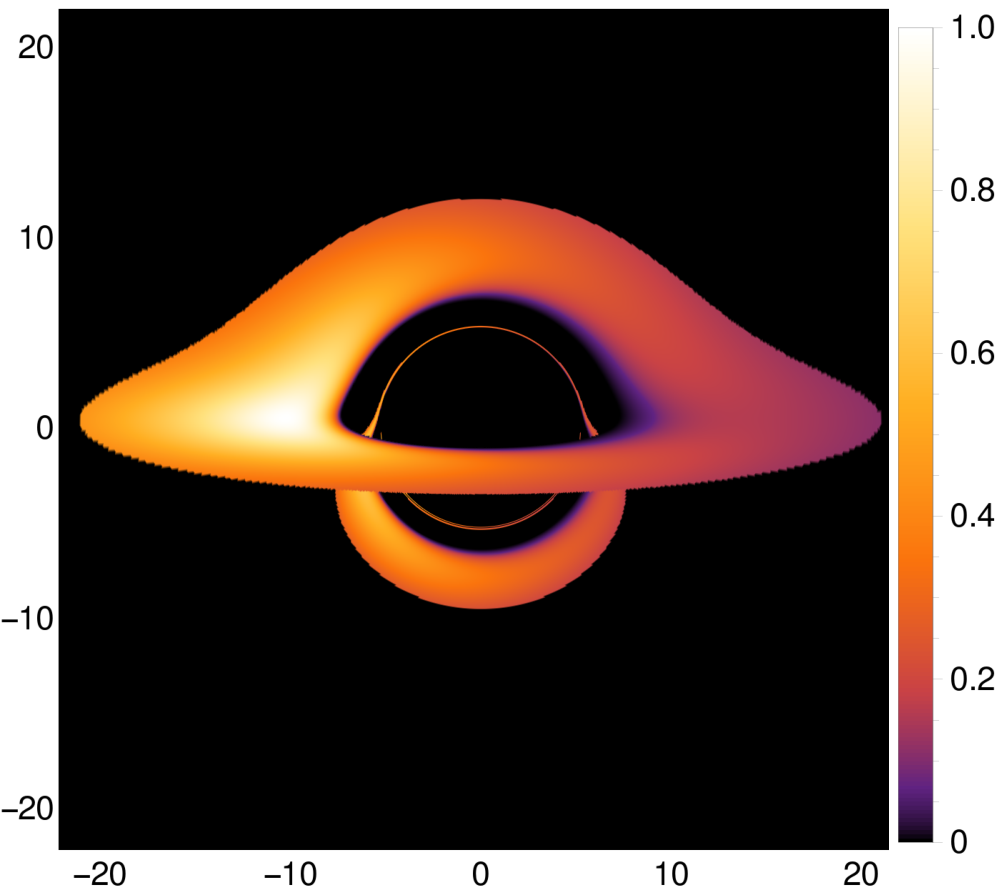}}
\subfigure[~$\gamma=0.75$]{\includegraphics[scale=0.54]{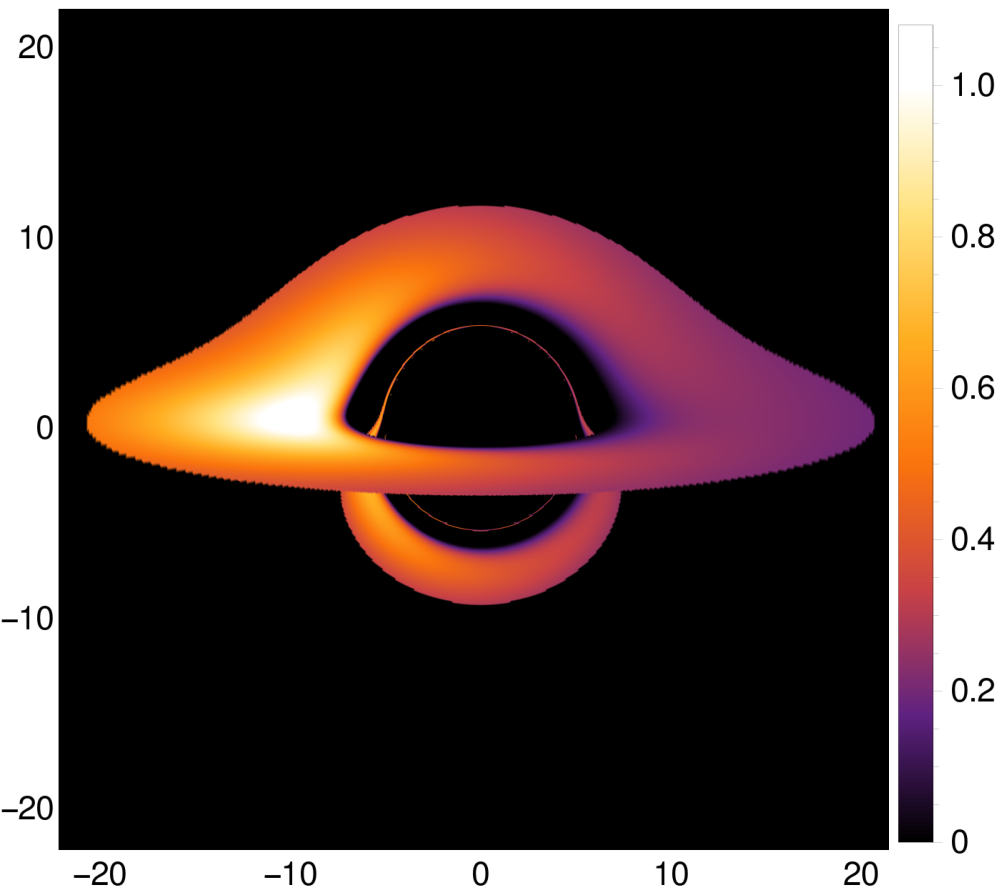}}
\subfigure[~$\gamma=0.51$]{\includegraphics[scale=0.54]{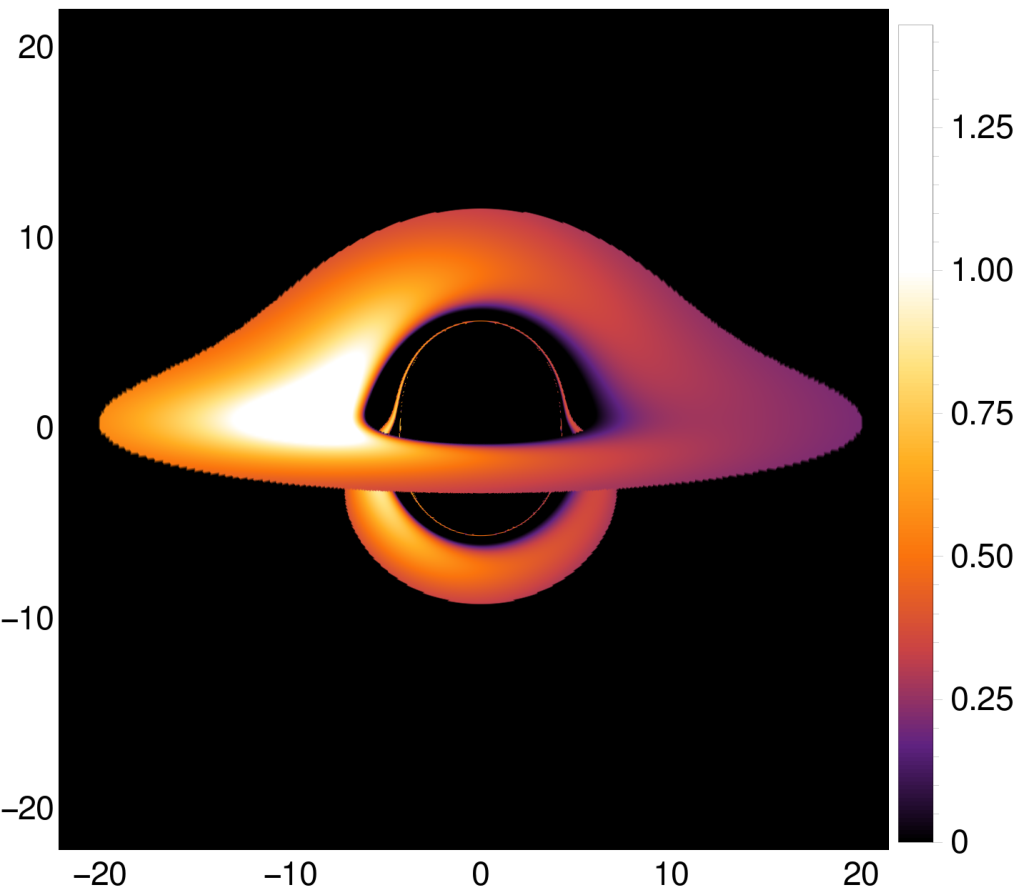}}
\subfigure[~$\gamma=0.45$]{\includegraphics[scale=0.46]{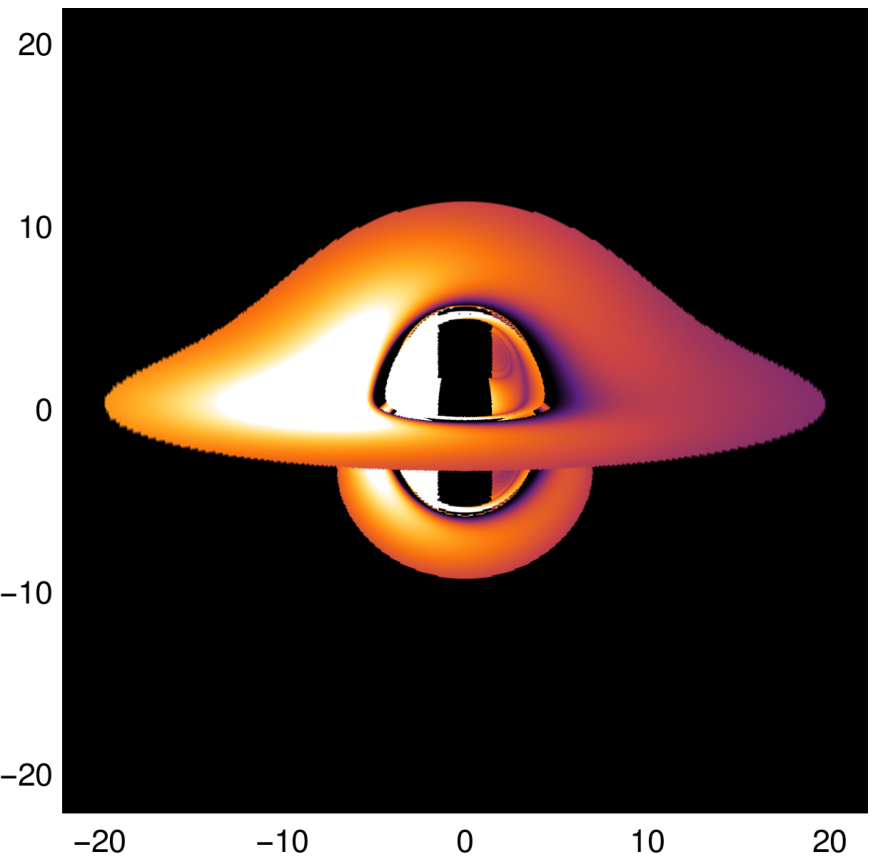}}
\subfigure[~$\gamma=0.4$]{\includegraphics[scale=0.46]{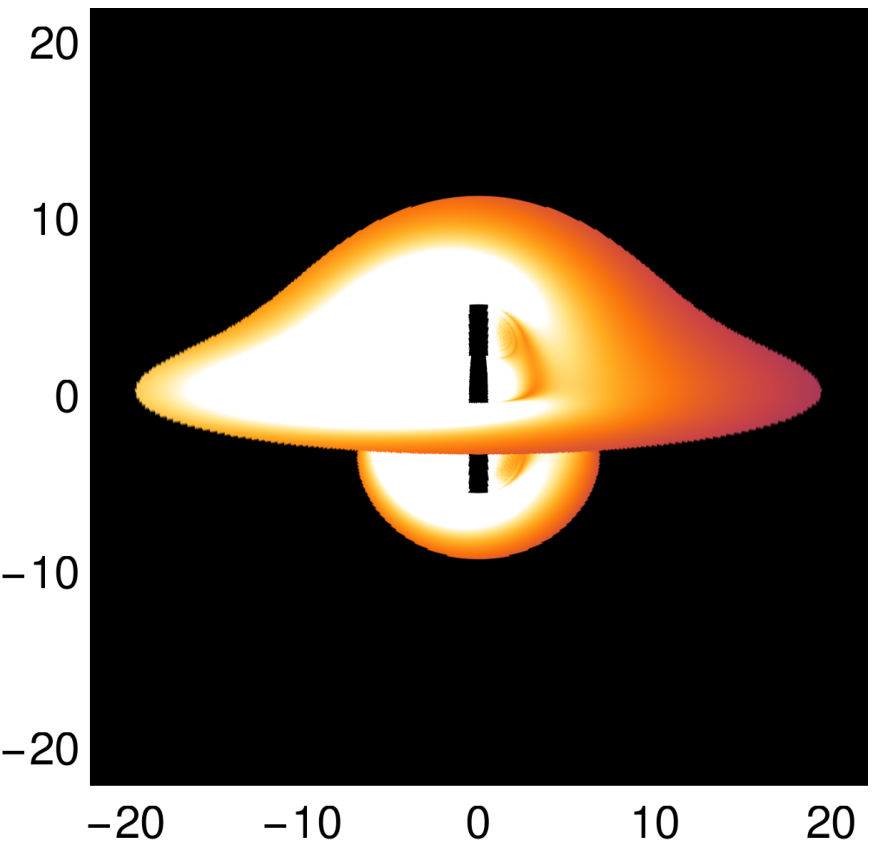}}
\subfigure[~$\gamma=0.3$]{\includegraphics[scale=0.46]{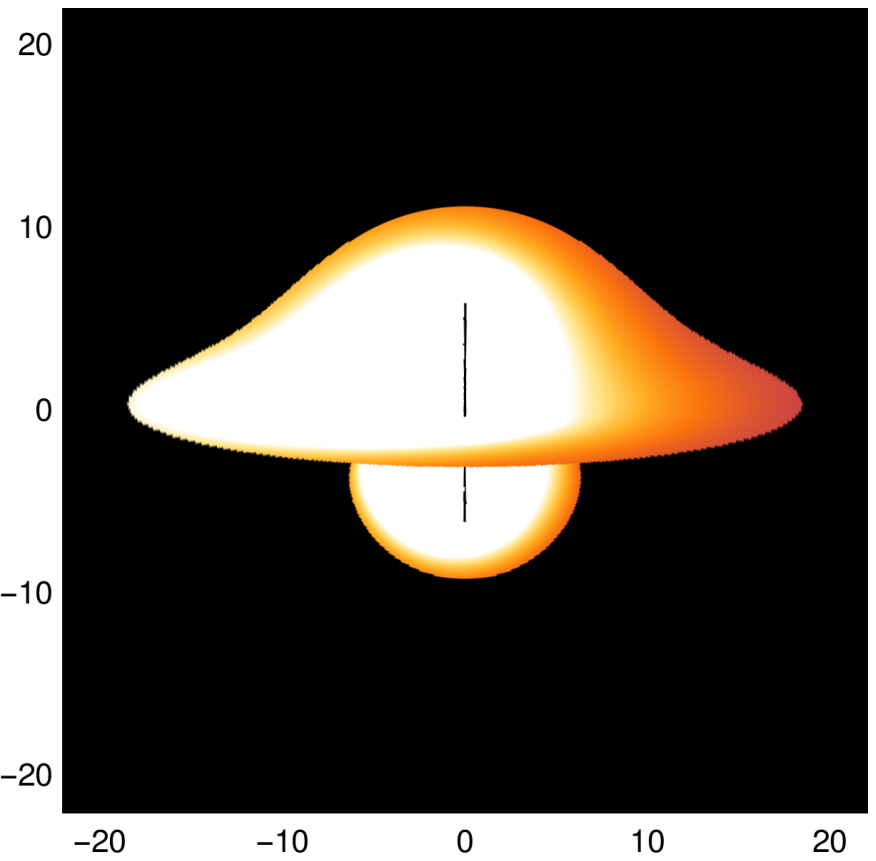}}
\subfigure[~$\gamma\to \infty$ (GI)]{\includegraphics[scale=0.445]{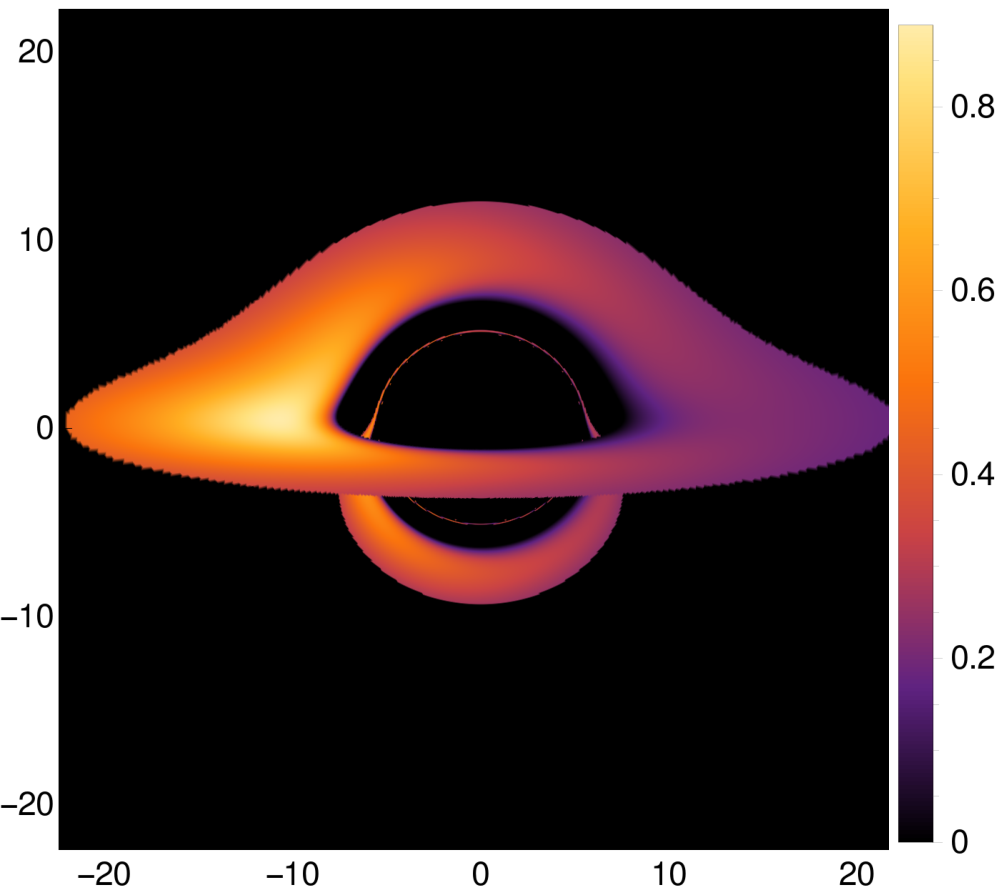}}
\caption{The images of accretion disks in a Schwarzschild black hole and the  $\gamma$-metric [(a)-(g)]. See the 
text for the disk configurations for different values of $\gamma$. The outer edge of the outer disk is at $r = 20M$, 
and the observer's inclination angle is $\theta_{o}=80^{\circ}$. The observer is placed at the radial coordinate 
$r=10^4M$, which corresponds effectively to the asymptotic infinity. In order to get rid of the parameters $M$ 
and $\dot{M}$, we have normalized the fluxes by the maximum flux observed for the Schwarzschild black hole. 
Also, we have plotted square-root of the normalized flux for better looking. The color bars show the values the flux. 
All spatial coordinates are in units of $M$. Here, we have taken $\delta r=10^{-9}$ for $\gamma< 1/2$ case. 
The central black strip in this case will disappear in the limit $\delta r\to 0$.}
\label{fig:disk_image}
\end{figure}

We now use our numerical ray-tracing techniques (with certain modifications) discussed in 
our previous work \cite{Suvankar_2020} (see also \cite{rajibul_disk}) and produce the images of accretion disks. The intensity maps of the images of accretion disks for the different cases discussed above are shown in 
Fig. \ref{fig:disk_image}. Note that, when there are light rings in the $\gamma$-metric, the image is very 
similar to that of a black hole, thereby mimicking a black hole. In the absence of light rings, however, 
the images for the $\gamma$-metric differ strikingly from that of the black hole, as shown by Figs. \ref{fig:disk_image}(d), \ref{fig:disk_image}(e) and \ref{fig:disk_image}(f).

\section{Conclusions}
\label{conclusions}

The unprecedented advances in observational studies in the current era of the EHT have resulted in the exciting
possibility of understanding the nature of singularities in GR, by comparing theory with experimental data. 
Motivated by the celebrated work of Hawking \cite{Hawking},
the importance of such studies lies in the fact that these can ultimately throw light on the nature of strong gravity, 
effective at horizon scales \cite{Giddings}. An important aspect of such analyses is the fact that one can use
the EHT data to constrain possible solutions of Einstein equations. Indeed, a plethora of activities have been
reported in this direction in the recent past, and it is by now understood that several singular solutions, as well as horizonless
compact objects are consistent with current data. The totality of such results will be pivotal in understanding the
correct nature of the fundamental aspects of strong gravity. 

In this paper, we have carried out the analysis of shadows and accretion disk images of the  Zipoy-Voorhees spacetime, 
characterized by the $\gamma$-metric. We have shown that while for the parameter $\gamma <1/2$, there will
be no shadow, the  $\gamma \geq 1/2$ class is essentially unconstrained, i.e., all such values of $\gamma$ are consistent
with current EHT observations. We have further constructed thin accretion disk images in the $\gamma$ metric background
and shown that these can be dramatically different from the Schwarzschild case for $\gamma<1/2$. 

As we mentioned in the introduction, astrophysical black holes are often approximated by the static Schwarzschild
or the stationary Kerr solution. The $\gamma$-metric on the other hand describes a static, axially-symmetric 
vacuum solution, and is attractive in its own rights. Since spherical symmetry in static vacuum solutions is by no means
a fundamental criterion in black hole physics, our result that the $\gamma$-metric is perfectly admissible,
should be an interesting addition to the current literature. Further, we have shown how the thin accretion disk
images here might be very similar to that of the Schwarzschild black hole in cases when there are light rings. These cases thus exemplify situations where the horizonless object might mimic a black hole. 
In the absence of any light ring, however, these images might be very different from the black hole case. 

In continuation of this work, it should be interesting to compare other static axially symmetric solutions of
GR with the current data from the EHT.


\begin{thebibliography}{99}

\bibitem{EHT1} The Event Horizon Telescope Collaboration, {\em First M87 event horizon telescope results. I. 
The shadow of the supermassive black hole}, Astrophys. J. Lett. {\bf 875}, L1 (2019).

\bibitem{EHT2} The Event Horizon Telescope Collaboration, {\em First M87 event horizon telescope results. V. 
Physical origin of the asymmetric ring}, Astrophys. J. Lett. {\bf 875}, L5 (2019).

\bibitem{EHT3} The Event Horizon Telescope Collaboration, {\em First M87 event horizon telescope results. VI. 
The shadow and mass of the central black hole}, Astrophys. J. Lett. {\bf 875}, L6 (2019).

\bibitem{tapo1} 
  R.~Shaikh, K.~Pal, K.~Pal and T.~Sarkar,
  {\em Constraining alternatives to the Kerr black hole},
  arXiv:2102.04299 [gr-qc].
  
\bibitem{Psaltis} D. Psaltis et al. (Event Horizon Telescope), {\em Gravitational Test Beyond the First Post-Newtonian Order with the Shadow of the M87 Black Hole}, Phys. Rev. Lett. {\bf 125}, 141104 (2020).

\bibitem{Soumitra} I.~Banerjee, S.~Chakraborty and S.~SenGupta,
{\em Silhouette of M87*: A New Window to Peek into the World of Hidden Dimensions},
  Phys.\ Rev.\ D {\bf 101}, no. 4, 041301 (2020).

\bibitem{Bambi} C. Bambi, K. Freese, S. Vagnozzi and L. Visinelli, {\em Testing the rotational nature of the supermassive object M87$^*$
from the circularity and size of its first image}, Phys. Rev. D {\bf 100}, 044057 (2019).
  
\bibitem{Rahul} R. Kumar and S. G. Ghosh, {\em Black Hole Parameter Estimation from Its Shadow}, ApJ {\bf 892}, 78 (2020).

\bibitem{Erez-Rosen_1959} G. Erez and N. Rosen, {\em The gravitational field of a particle possessing a multipole moment}, 
Bull. Res. Council Israel 8, 47 (1959).

\bibitem{Zipoy_1966} D. M. Zipoy, {\em Topology of some spheroidal metrics}, J. Math. Phys. 7, 1137 (1966).

\bibitem{Voorhees_1970} B. H. Voorhees, {\em Static axially symmetric gravitational fields}, Phys. Rev. D {\bf 2} 2119 (1970).

\bibitem{Esposito_1975} F. P. Esposito, L. Witten, {\em On a static axisymmetric solution of the Einstein equations}, 
Phys. Lett. {\bf B58}, 357 (1975).

\bibitem{Gutsunaev_1985} Ts. I. Gutsunaev, V. S. Manko, {\em On the Gravitational Field of a Mass
Possessing a Multipole Moment}, Gen. Rel. Grav. {\bf 17} 1025 (1985),

\bibitem{Stephani} H. Stephani, D. Kramer, M. Maccallum, C. Hoenselaers, E. Herlt, {\em Exact
Solutions to Einstein's Field Equations} 2 Ed., Cambridge University Press (2003). 

\bibitem{Griffiths_2009} J. B. Griffiths and J. Podolsky, {\em Exact Space-Times in Einstein's General 
Relativity}, Cambridge University Press, Cambridge, United Kingdom, 2009.

\bibitem{Herrera_1999} L. Herrera, F. M. Paiva, N. O. Santos, {\em The Levi-Civita space–time as a limiting 
case of the $\gamma$ space–time}, J. Math. Phys. 40, 4064 (1999).

\bibitem{Kodama_2003} H. Kodama and W. Hikida, {\em Global structure of the Zipoy-Voorhees-Weyl 
spacetime and the delta=2 Tomimatsu-Sato spacetime}, Classical and Quantum Gravity 20, 5121 (2003).

\bibitem{Galtsov_2019} D. V. Gal'tsov, K. V. Kobialko, {\em Photon trapping in static axially symmetric spacetime}, 
Phys. Rev. D {\bf 100} 104005 (2019).

\bibitem{Bambi_2019} A. B. Abdikamalov, A. A. Abdujabbarov, D. Ayzenberg, D. Malafarina, C. Bambi, and 
B. Ahmedov, {\em A black hole mimicker hiding in the shadow: Optical properties of the $\gamma$-metric}, 
Phys. Rev. D {\bf 100}, 024014 (2019).

\bibitem{Herrera_2005} L.~Herrera, G.~Magli and D.~Malafarina,
 ``Non-spherical sources of static gravitational fields: Investigating the boundaries of the no-hair theorem,''
 Gen.\ Rel.\ Grav.\  {\bf 37}, 1371 (2005)

\bibitem{Virbhadra_1996} K. S. Virbhadra, {\em Directional naked singularity in general relativity}, arXiv:gr-qc/9606004.

\bibitem{Boshkayev_2016} K. Boshkayev, E. Gasperin, A. C. Gutierrez-Pineres, H. Quevedo, and S. Toktarbay, 
{\em Motion of test particles in the field of a naked singularity}, Phys. Rev. D 93, 024024 (2016).

\bibitem{Quevedo_2011} H. Quevedo, {\em Mass quadrupole as a source of naked singularities}, 
Int. J. Mod. Phys. D 20, 1779 (2011).

\bibitem{Hernandez_1994} J. L. Hernández-Pastora and J. Martín, 
{\em Monopole-Quadrupole Static Axisymmetric Solutions of Einstein Field Equations}, Gen. Rel. and Grav. 26, 877 (1994).

\bibitem{Chowdhury_2012} A. N. Chowdhury, M. Patil, D. Malafarina, and P. S. Joshi, {\em Circular 
geodesics and accretion disks in the Janis-Newman-Winicour and gamma metric spacetimes}, Phys. Rev. D 85, 104031 (2012).

\bibitem{Chazy_1924} J. Chazy, {\em Sur la champ de gravitation de deux masses fixes dans la theory 
de la relativit'e}, Bull. Soc. Math. Fr. 52, 17 (1924).

\bibitem{Curzon_1924} H. E. J. Curzon, {\em Cylindrical solutions of Einstein’s gravitation equations}, 
Proc. London Math. Soc. 23, 477 (1924).

\bibitem{Camacho_2015} P. M. Camacho, F. F. Alfaro, and C. G. Chaves, {\em Slowly rotating Curzon-Chazy metric}, 
Rev. Mat. Teor. Apl. 22, 265 (2015).

\bibitem{Suvankar_2020} S. Paul, R. Shaikh, P. Banerjee and T. Sarkar, {\em Observational signatures of 
wormholes with thin accretion disks}, JCAP 03 (2020) 055.

\bibitem{novikov_1973} I. D. Novikov and K. S. Thorne, {\em Astrophysics and black holes}, in {\em Black Holes}, 
edited by C. DeWitt and B. DeWitt (Gordon and Breach, New York, 1973).

\bibitem{page_1974} D. N. Page and K. S. Thorne, {\em Disk-accretion onto a black hole. Time-averaged structure 
of accretion disk}, Astrophys. J. {\bf 191}, 499 (1974).

\bibitem{rajibul_disk} R. Shaikh and P. S. Joshi, {\em Can we distinguish black holes from naked singularities by the images of their accretion disks?}, JCAP 10 (2019) 064.

\bibitem{Hawking}
S.~W.~Hawking,
{\em Particle Creation by Black Holes},
Commun.\ Math.\ Phys.\  {\bf 43}, 199 (1975)
Erratum: [Commun.\ Math.\ Phys.\  {\bf 46}, 206 (1976)].

\bibitem{Giddings}
S.~B.~Giddings,
{\em Searching for quantum black hole structure with the Event Horizon Telescope},
Universe {\bf 5}, no. 9, 201 (2019)

\end{thebibliography}
\end{document}